\documentclass[11pt]{article}
\usepackage[utf8]{inputenc}
\usepackage[T1]{fontenc}\usepackage{lmodern}   %
\usepackage[margin=1in]{geometry}
\usepackage{graphicx,booktabs,xcolor}
\usepackage{placeins}
\usepackage[numbers,sort&compress]{natbib}
\usepackage[colorlinks=true,linkcolor=blue,citecolor=blue,urlcolor=blue]{hyperref}
\usepackage{xurl}  %
\usepackage{tikz}\usetikzlibrary{arrows.meta,positioning,calc}
\newcommand{\tnr}{\textmd{\textsc{nr}}}\newcommand{\tfeat}{\textmd{\textsc{feat}}}
\newcommand{\tpseudo}{\textmd{\textsc{pseudo}}}\newcommand{\tstrfn}{\textmd{\textsc{strfn}}}
\title{General-purpose language models are neurosymbolic reasoners able to annotate proteins from sequence}
\author{Christopher J. Mungall\thanks{Corresponding author: \href{mailto:cjmungall@lbl.gov}{cjmungall@lbl.gov}} \qquad Marcin P. Joachimiak\\[1ex]
\small Environmental Genomics and Systems Biology Division,\\
\small Lawrence Berkeley National Laboratory, Berkeley, CA 94720, USA}
\date{}

\begin{document}
\maketitle

\begin{abstract}
  Determining the function of genes across the tree of life is one of the major challenges in biology. Generative AI approaches
  to this problem generally fall into two categories: language models trained on biological sequences that are able to predict
  structure and function from signals in those sequences; and generalist language models trained on text, that are able to reason
  about biological function symbolically. A default operating assumption is that protein language models do not perform well on symbolic tasks,
  and that generalist language models do not perform well on sequence tasks. However, a recent report showed that frontier generalist models
  perform surprisingly well on next-residue prediction tasks. We set out to explore whether
  these capabilities extended to protein annotation.

  We assigned different frontier generalist models a variety of protein annotation tasks, including predicting the
  presence and coordinates of Pfam domains, loss of functional catalytic residues, and predicting structure and activity. On
  all annotation tasks, the frontier models performed well above chance; we did not benchmark them against specialised tools.
  The models also make characteristic hallucination errors, and they state them as confidently as their correct calls.

  Though untested against specialised methods, the nascent ability of frontier models to perform protein annotation from sequence alone
  is intriguing. The ability to reason over sequence, structure, and symbols in a single model allows for a neurosymbolic approach where
  a combination of sequence, experimental data and knowledge from the literature can be holistically combined to improve on de novo functional annotation of the protein universe.

Prompts, test sets, model outputs and code:\newline
\url{https://github.com/ai4curation/fable-sequence-to-function}\newline
Summary of results: \url{https://ai4curation.io/fable-sequence-to-function/}
\end{abstract}

\section{Introduction}

Protein language models such as ESM-2 and ProtGPT2 learn sequence statistics from protein databases and
support structure prediction and design \citep{lin2023esm2,ferruz2022protgpt2}. General-purpose language
models are built for other tasks. Even so, Tu and Chen \citep{tu2026claude,tu2026probe} recently reported that two of them carry sequence
knowledge. On proteins from little-studied domain-of-unknown-function families, Claude Fable~5.2 and
Claude Opus~5.5 predicted the next amino acid much more often for real sequences than for shuffled ones,
at a level comparable to dedicated protein language models. The same report found little or no such
signal in the models it tested from other vendors, and a smaller one in earlier Claude models. We tested four models from two providers: those two, an earlier Claude
model and a Gemini model.

General-purpose models have also been benchmarked on ranking protein variants. PG-LLM gave thirteen such
models a wild-type sequence, a description of the assay and a list of variant sequences to rank, in 276
tasks, 217 of them from ProteinGym and 59 from recently published studies, with no tools, alignment or structure. The best of them outperformed most
sequence-only predictors and stayed below the leading specialist predictor \citep{arora2026pgllm}. That
benchmark scores a ranking of whole sequences against measured fitness. The tests here ask instead for
explicit annotation of residues and features, with matched decoys, and follow two disagreements with the
curated record through to the evidence.

Next-residue accuracy does not say whether a model can do anything a curator needs. Function prediction
tools make characteristic errors that human annotators avoid \citep{decrecy2025limitations}. One that
matters for enzymes is assigning a catalytic activity to a protein that has kept the fold of an enzyme
but lost the activity. Such proteins are called pseudoenzymes, or dead enzymes
\citep{murphy2017biozombie,adrain2020dead}. A kingdom-wide analysis estimated that they make up 5 to 10\%
of the proteins in enzyme families \citep{ribeiro2019pseudoenzymes}, and a review of their annotation in
UniProtKB gives 10 to 15\% \citep{zaru2020uniprot}. Many act as regulators \citep{adrain2020dead}, and
pseudokinases are the best-studied case \citep{kwon2019pseudokinases}. A model that only recognises the
family will repeat the error; a model that reads the residues should not.

Reading the residues has limits too. Losing a catalytic residue predicts inactivity but does not prove
it. A survey of enzyme families found that mutation of catalytic residues is not always a good indication
of lack of activity \citep{ribeiro2020pitfalls}. In some proteins the residues needed for catalysis have
moved elsewhere in the sequence \citep{goldberg2023pseudoenzymes}. WNK1, for example, lacks the catalytic
lysine of kinase subdomain~II yet is an active kinase, and its activity requires Lys233, in
subdomain~I \citep{xu2000wnk1}.

We ran four small tests, with controls meant to show whether the models respond to the sequence itself:
catalytic-activity calls on kinases and other enzymes carrying a single
substitution that we introduced; domain and membrane-feature annotation, with shuffled
sequences as the control; secondary structure, contact and function prediction for recently
released structures, stratified by prior structural homology; and a subsample replication of the
next-residue result.

\section{Results}

\subsection{Study design}

\begin{figure}[t]
\centering
\begin{tikzpicture}[
  font=\footnotesize, line width=0.6pt, >={Stealth[length=1.8mm,width=1.4mm]},
  every node/.append style={execute at begin node={\hyphenpenalty=10000\exhyphenpenalty=10000\relax}},
  box/.style={draw, rounded corners=2pt, align=left, inner sep=4pt, text width=#1},
  tag/.style={draw, rounded corners=2pt, fill=black!7, minimum width=1.5cm, minimum height=0.6cm, inner sep=2pt, font=\small},
  cell/.style={align=left, text width=#1, inner sep=2pt, anchor=west},
  head/.style={font=\footnotesize\bfseries, anchor=west, inner sep=2pt},
]
\node[box=2.0cm] (in) at (1.14,0) {%
  \textbf{Input}\\[2pt]
  an amino-acid sequence, nothing else\\[4pt]
  {\ttfamily\scriptsize ..LVHGDVKLQ..}};
\node[box=2.6cm, right=0.45cm of in] (llm) {%
  \textbf{Language model}\\[2pt]
  Frontier: Fable~5.2, Opus~5.5, Gemini~4 Neon\\[2pt]
  Earlier: Opus~4.6\\[4pt]
  {\scriptsize no alignment or search requested}};
\draw[->] (in) -- (llm);
\foreach \y/\code/\task/\ctrl/\score [count=\i] in {
  2.25/\tpseudo/{Active or not? Which catalytic residues are present or substituted?}/{wild type\\ catalytic mutant\\ matched decoy}/{curated catalytic sites},
  0.75/\tfeat/{Pfam domains and their boundaries; transmembrane helices; signal peptide}/{same protein, shuffled}/{InterPro and UniProt features},
  -0.75/\tstrfn/{Secondary structure, long-range contacts, function}/{chance; best single-state guess}/{PDB entries released after Aug 2026},
  -2.25/\tnr/{Next residue after a cut point}/{same protein, shuffled}/{the true next residue}%
}{
  \node[tag] (t\i) at (6.85,\y) {\code};
  \node[cell=3.4cm] at (7.75,\y) {\task};
  \node[cell=2.6cm] at (11.4,\y) {\ctrl};
  \node[cell=1.95cm] at (14.25,\y) {\score};
  \draw[->] (llm.east) -- ++(0.22,0) |- (t\i.west);
}
\node[head] at (6.1,3.3) {Task (what the model returns)};
\node[head] at (11.4,3.3) {Control};
\node[head] at (14.25,3.3) {Reference};
\draw[black!40, line width=0.4pt] (6.1,3.0) -- (16.3,3.0);
\foreach \y in {1.5,0,-1.5} \draw[black!15, line width=0.4pt] (6.1,\y) -- (16.3,\y);
\end{tikzpicture}
\caption{Study overview. Each model receives an amino-acid sequence and one of four tasks, named by the
short codes used throughout. Every task has a control meant to separate reading the sequence from
recalling a family label.}
\label{fig:overview}
\end{figure}
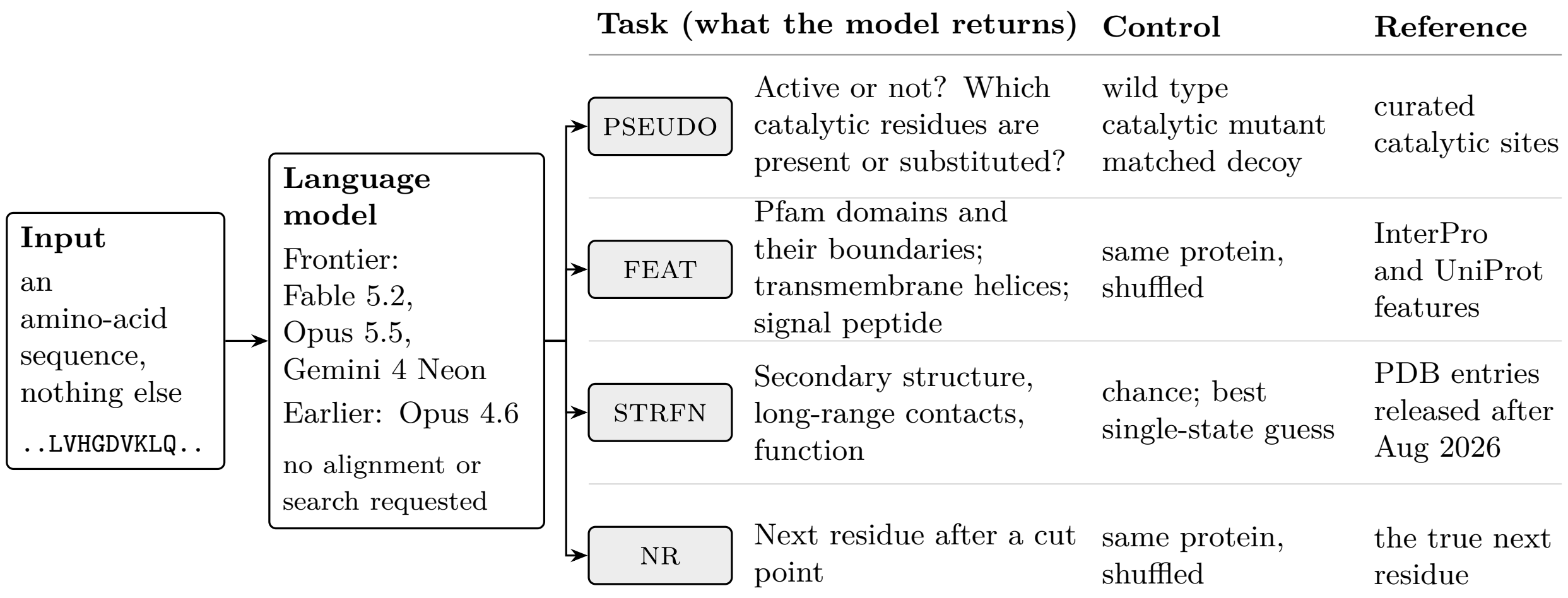

Figure~\ref{fig:overview} summarises the design. We tested four models from two providers (Table~\ref{tab:models}), each on all four tests:
\texttt{claude-fable-5-2}, \texttt{claude-opus-5-5}, \texttt{claude-opus-4-6} and
\texttt{gemini-4-neon}. All models received the same frozen test sets and the same
prompts. Effort settings are each provider's own labels and differ between models and tests. Because
they are not comparable reasoning budgets, we also report the thinking tokens each model recorded
(Table~\ref{tab:thinking}). Opus~4.6 was run at two settings in the \tfeat, \tpseudo{} and \tstrfn{} tests; unless stated, we report its
high-effort run. Model identifiers are those returned by the endpoints. We did not establish any model's
training cutoff. Runs of other Gemini models on some of the tests (Gemini~3.1 Pro Preview and Gemini~3.1 Flash Lite)
and incomplete runs of other routes are kept in the repository and are not reported here.

We call Fable~5.2, Opus~5.5 and Gemini~4 Neon the frontier models. Opus~4.6 is an earlier generation of
Claude model, and we use it as a contrast.

The runs differ in how well they exclude outside information, and Table~\ref{tab:routes} lists the
three routes. The first Claude route cannot rule out that some calls carried extra context or connector
tools. The \tfeat{} and \tstrfn{} results of Fable~5.2 and Opus~5.5 come from that route only, as do
their calls on the 20 proteins named inactive and, apart from eight Opus~5.5 questions, their \tnr{}
answers. Of their results, only the \tpseudo{} wild types and mutants were repeated by the revised route. Details
of retries, input sizes and software versions are in the Supplementary Methods.

\begin{table}[ht]
\centering\small
\caption{How the models were called. We supplied no tools and no system prompt on any route.}
\label{tab:routes}
\begin{tabular}{p{0.2\linewidth}p{0.36\linewidth}p{0.36\linewidth}}
\toprule
Route & What it excludes & Runs \\
\midrule
Claude command-line tool, original flags (18 November) & Built-in tools off, empty system prompt. Extra context or connector tools in a call cannot be ruled out. & Fable~5.2 and Opus~5.5: all four tests. Opus~4.6: all four tests at low effort. \\
\addlinespace
Claude command-line tool, revised flags (19 November) & As above, and user settings and connector tools are kept out of the call. & Opus~4.6 at high effort (\tpseudo, \tfeat, \tstrfn). Decoy control, with a repeat of the \tpseudo{} wild types and mutants. Eight Opus~5.5 \tnr{} questions left unasked by the first run. \\
\addlinespace
Vertex AI API (19 November) & Request body carries no tools and no system instruction. & Gemini~4 Neon: all four tests. \\
\bottomrule
\end{tabular}
\end{table}

\begin{table}[ht]
\centering\small
\caption{Models and the effort setting of each reported run. Labels are the providers' own and are not matched across providers.}
\label{tab:models}
\begin{tabular}{llllll}
\toprule
Model & Provider & \tpseudo & \tfeat & \tstrfn & \tnr \\
\midrule
Claude Fable 5.2 & Anthropic & high & high & high & low, high \\
Claude Opus 5.5 & Anthropic & high & high & high & low, high \\
Claude Opus 4.6 & Anthropic & low, high & low, high & low, high & low \\
Gemini 4 Neon & Google & low & low & medium & low \\
\bottomrule
\end{tabular}
\end{table}

Each model was given a protein sequence and asked one of four things (details in
Section~\ref{sec:methods}). We refer to the tests by the short names in parentheses. The prompts, test sets and every
model reply are in the project repository (\url{https://github.com/ai4curation/fable-sequence-to-function}).

\textbf{Catalytic residues and pseudoenzymes (\tpseudo).} Say whether an enzyme is catalytically active and which catalytic
residues are present or substituted. The sequences were 20 active kinases, the same kinases with the
catalytic aspartate or the ATP-binding lysine substituted, 20 kinase-fold proteins that Swiss-Prot names
``inactive'', and 20 other enzymes with and without an active-site residue changed to alanine. In a decoy
control, run for the Claude models, the same type of substitution was placed at a position with no
functional annotation. What the test measures is sensitivity: whether a model's call changes when an
annotated catalytic residue is substituted, and not when a matched residue elsewhere is. We did not
measure the activity of any mutant or decoy.

\textbf{Domains and membrane features (\tfeat).} Name the Pfam domains with their boundaries, the
transmembrane helices and any signal peptide, for 45 sampled bacterial and archaeal proteins and 15 from
obscure domain-of-unknown-function families. Each protein was also given in shuffled form.

\textbf{Structure and function (\tstrfn).} Predict secondary structure, residue contacts and function for
30 PDB structures released from August 2026. We call an entry ``unmatched'' when no PDB entity at 30\%
sequence identity or more was released before June 2026, and ``matched'' when at least 20 were. The labels
describe that search and nothing else: an unmatched protein is not necessarily new to science or to the
models. One unmatched protein, human ASPRV1, was drawn four times and is counted once, leaving 12
unmatched proteins.

\textbf{Next residue (\tnr).} Continue a sequence from a cut point. The control is the same protein with
its residues shuffled.

\subsection{A single catalytic substitution is detected, and a matched decoy is ignored}

\begin{figure}[ht]
\centering
\includegraphics[width=\linewidth]{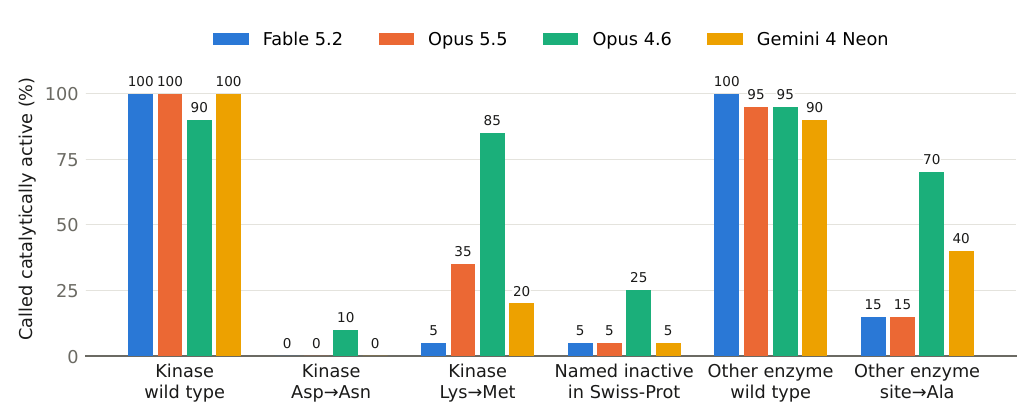}
\caption{Share of sequences called catalytically active, 20 sequences per bar. Mutants carry one
substitution introduced for this study. Claude models at high effort; Gemini~4 Neon at low effort.}
\label{fig:pseudo}
\end{figure}

Fable~5.2, Opus~5.5 and Neon each called all 20 wild-type kinases active and none of the 20
aspartate-to-asparagine mutants (Figure~\ref{fig:pseudo}). Across the 60 mutant sequences, they reported
the mutated position as substituted in 57, 47 and 47 (Table~\ref{tab:positions}). Opus~5.5 and Neon
changed their call less often for the lysine: they called 7 and 4 of the 20 lysine-to-methionine mutants active. Neon
also called 8 of 20 alanine mutants of the other enzymes active. An ``active'' call on a mutant goes
against our prediction that the mutant is inactive. We did not test that prediction, so such a call is
not a demonstrated error.

Opus~4.6 changed its call far less often for the lysine and the other enzymes. At high effort, it called 2 of 20 aspartate-to-asparagine
mutants active and named the substituted position in 18. But it called 17 of 20 lysine-to-methionine
mutants and 14 of 20 mutants of other enzymes active, and named the position in only 1 and 2 of them. At
low effort it called 18, 18 and 16 of 20 active and named no position, so its result on the aspartate
depends on effort.

\begin{table}[ht]
\centering\small
\caption{The \tpseudo{} test: mutants in which the model reported the substituted position, of 20 per row. Claude models
at high effort; Neon at low effort.}
\label{tab:positions}
\begin{tabular}{lrrrr}
\toprule
Mutant class & Fable 5.2 & Opus 5.5 & Opus 4.6 & Neon \\
\midrule
Kinase Asp$\to$Asn & 20 & 19 & 18 & 20 \\
Kinase Lys$\to$Met & 20 & 13 & 1 & 16 \\
Other enzyme site$\to$Ala & 17 & 15 & 2 & 11 \\
\bottomrule
\end{tabular}
\end{table}

The decoy control shows that Fable~5.2 and Opus~5.5 respond to the catalytic position and not to just
any change (Table~\ref{tab:decoy}). With every question asked by one route, both models called all 58
decoys active, while calling only 3 and 8 of the 58 catalytic-site mutants of the same proteins active.
Neither named a decoy position as a lost residue. One case ran the other way: Opus~5.5 did not call the
wild type of a mannosyltransferase (P26725) active, but called its decoy active. Opus~4.6, given the
decoys at low effort only, called decoys, mutants and wild types active at similar rates.

Asking the mutants and wild types again by the revised route also gave a replicate of the \tpseudo{} test. The
active-or-not call matched the first run for 99 of 100 questions for Fable~5.2 and 96 of 100 for
Opus~5.5. We ran no decoy control for Neon. It called all 20 wild-type kinases active,
which rules out a blanket ``inactive'' answer but not a tendency to reject any altered sequence.

\begin{table}[ht]
\centering\small
\caption{Decoy control, all columns asked by one route: 38 proteins, 58 catalytic-site mutants and 58
decoys carrying the same substitution type at an unannotated position. Counts are sequences called
catalytically active.}
\label{tab:decoy}
\begin{tabular}{lrrr}
\toprule
Model (effort) & Wild type & Catalytic-site mutant & Decoy \\
\midrule
Fable 5.2 (high) & 38/38 & 3/58 & 58/58 \\
Opus 5.5 (high) & 37/38 & 8/58 & 58/58 \\
Opus 4.6 (low) & 34/38 & 45/58 & 50/58 \\
\bottomrule
\end{tabular}
\end{table}

Fable~5.2, Opus~5.5 and Neon called 18, 18 and 19 of the 20 proteins that Swiss-Prot names inactive
kinases inactive. This is agreement with a database name, not a measure of biological accuracy. In 17, 17
and 19 of those answers the protein name the model gave already said ``inactive'' or ``pseudokinase'', so
the set mostly measures recall of the entry. Nor does the name establish inactivity: we selected the 20
by name and checked only three of the labels against the literature (below). Opus~4.6 at high effort called 12 of 20
inactive.

For at least two of the 20 the name is contradicted by a published assay. The kinase domains of
\textit{Arabidopsis} RLK902 (Q9LVI6) and RKL1 (Q9LP77), purified from \textit{E.~coli} as fusion
proteins, each became phosphorylated when incubated with radiolabelled ATP, and the authors concluded
that both are ``functional protein kinases capable of autophosphorylation'' \citep{tarutani2004rlk902}.
The Swiss-Prot entry for RLK902 notes this result beside its prediction that the kinase domain is inactive
\citep{uniprot}. Fable~5.2, Opus~5.5, Neon and Opus~4.6 at high effort called both proteins inactive.
Agreeing with the name is therefore not the same as being right. One in vitro autophosphorylation assay does
not settle what these proteins do in the plant either.

All three of the stronger models called \textit{C.~elegans} kin-32 (Q95YD4) active, although Swiss-Prot
names it an inactive tyrosine kinase. They cited an intact glycine loop, VAVK lysine, HRD and DFG motifs,
and all of these are present in the sequence. The Swiss-Prot name rests on an assay, not on the sequence:
the entry records that the protein ``has apparently no tyrosine kinase activity in vitro when expressed in
mammalian cells''. In that study, kin-32 immunoprecipitated from transfected mouse cells showed no
measurable activity against a general tyrosine kinase substrate \citep{cram2008kin32}. The authors note
that it lacks the tyrosine at which mammalian focal adhesion kinase autophosphorylates, and they allow
that it may be a kinase in the worm that was not activated in mouse cells. Intact canonical motifs
therefore do not guarantee detectable activity, and a reading
of catalytic residues cannot tell such a case from an active kinase. Nor does one assay settle whether
kin-32 is inactive in the worm.

Two patterns recurred among the mutants that were called active. In the first, the model moved the catalytic role to a neighbouring residue of
the same type. When we changed Glu318 of a glucuronosyltransferase (Q9VLA1) to alanine, Fable~5.2 and
Opus~5.5 named Glu317 as the catalytic base and called the enzyme active, although Fable~5.2 had named
Glu318 in the wild type. Neon also called this mutant active, but for a different reason: it listed
Asp233 and Asp235 and no catalytic glutamate in either the wild type or the mutant.

In the second pattern, the model reported the substitution and then argued around it. For the
\textit{Toxoplasma} kinase ROP17 (Q2PAY3), Fable~5.2 noted methionine at the lysine position and proposed
that a nearby lysine would take its place. A displaced catalytic lysine has a precedent in WNK1
\citep{xu2000wnk1}, but here the model applied the argument to a substitution we had made in an active
kinase. In this pattern the model did detect the substitution; what departs from our prediction is its
call on activity.

Two further disagreements turn on disputed reference positions. For the nucleotidase YrfG (P64636)
and the reductase ENV9 (Q08651), Fable~5.2, Opus~5.5 and Neon each placed the catalytic residue at a
different position from UniProt, in wild type and mutant alike, and so called the mutants active. We do
not count agreement among models as evidence for a site, and it is not used below. Both UniProt
positions are annotated by similarity, and neither protein has an experimental structure.

For YrfG, UniProt gives the nucleophile as Asp9 and the models give Asp16. We aligned 11 reviewed haloacid-dehalogenase-superfamily relatives. Seven of them carry an
experimentally supported nucleophile, and all seven map to Asp16. In two structural models, Asp16 lies within 3.3~\AA{} of the magnesium ion and 2.6~\AA{} of
Lys149, while Asp9 is more than 20~\AA{} from both. The evidence strongly favours Asp16.

For ENV9, UniProt gives the catalytic serine as Ser172 and the models give the adjacent Ser173. UniProt's
cited source, O93868, has its catalytic serine at Ser149, which aligns with Ser173. That assignment comes
from a crystal structure \citep{horer2001mannitol}, and UniProt records it as curator inference
(ECO:0000305). In both structural models, the side chain of Ser173 points at the hydroxyl of the
catalytic Tyr202 (3.2 and 4.5~\AA), while that of Ser172 points away (7.0 and 6.8~\AA). The ENV9 case is
still tentative, because the candidates are adjacent and the source assignment is inferred from
structure. The published activity assays of ENV9 mutated the predicted active-site Asn146 (N146L) and the
cofactor-binding glycines (G23--24A), and each mutation left only residual activity. That paper names
Tyr202 and Lys206 as active-site residues and neither mutates nor names a catalytic serine
\citep{siddiqah2017env9}.

These two cases cast doubt on whether our substitutions removed the intended catalytic residues. They do
not show that the mutants keep their activity. In a sensitivity analysis that excludes the two proteins,
Fable~5.2 and Opus~5.5 each called 1 of 18 other-enzyme mutants active and Neon 6, and they reported the
substituted position in 17, 15 and 11 of 18. Figure~\ref{fig:pseudo} keeps all 20.

\subsection{Three models recover domains and membrane features and rarely assign a domain to a shuffled sequence}

Fable~5.2, Opus~5.5 and Neon named at least one correct Pfam family for 39, 35 and 20 of the 45 sampled
proteins. Their boundaries were close to the reference, with a mean overlap of 0.85, 0.85 and 0.80 over
the correctly named domains in these proteins (Table~\ref{tab:feat} pools all 60). For the 15 proteins
from obscure domain-of-unknown-function families, Fable~5.2 named the exact family for 8, Opus~5.5 for 3
and Neon for none. Neon was the most conservative. It predicted 28 domains, 23 with the right accession,
against 70 and 51 for the other two. All three almost never
assigned a Pfam domain to a shuffled sequence (1, 0 and 0 of 60). They were not left unannotated,
though: the three models predicted a transmembrane helix in 12, 15 and 15 of the 60. That need not be an
error, because a shuffled sequence keeps its composition and can still contain a hydrophobic stretch.

Opus~4.6 at high effort named a correct family for 7 of the 45 sampled proteins and assigned a domain to
28 of the 60 shuffled sequences. Its earlier run gave 5 of 45 and 5 of 60. That run was at low effort by
the authors' record; the \tfeat{} rows of 18 November do not store the setting. The high-effort run, on the
revised route, thus invented many more domains for little gain in recall.

The models gave no sign when they were wrong. Fable~5.2 assigned one DUF968 protein to a different
family. All three models split helices in multi-pass transporters, predicting 85, 87 and 82 helices
against 74 annotated, so recovering every reference helix does not mean the membrane annotation is right.
Two test proteins carry implausible submitter names in UniProt: ``Glycerol dehydrogenase'' on a
128-residue DUF3119 membrane protein, and ``Methylthioribose kinase'' on a DUF7147 protein. Neither
Fable~5.2 nor Opus~5.5 repeated them.

\begin{table}[ht]
\centering\small
\caption{Annotation from sequence alone. IoU is the overlap of predicted and reference domain boundaries
when the family is correct. The first two rows count proteins in each set; the other rows pool both sets
(60 proteins). Claude models at high effort, Neon at low effort. The stored Fable~5.2 and Opus~5.5 rows do
not record the setting (Methods).}
\label{tab:feat}
\setlength{\tabcolsep}{4pt}
\begin{tabular}{lrrrr}
\toprule
 & Fable 5.2 & Opus 5.5 & Opus 4.6 & Neon \\
\midrule
Sampled proteins: correct Pfam family & 39/45 & 35/45 & 7/45 & 20/45 \\
Obscure DUFs: correct Pfam family & 8/15 & 3/15 & 0/15 & 0/15 \\
Predicted domains: right accession & 62/70 & 45/51 & 9/40 & 23/28 \\
Boundary overlap (IoU) & 0.86 & 0.86 & 0.64 & 0.80 \\
Transmembrane helices recovered & 74/74 & 74/74 & 67/74 & 74/74 \\
Signal peptide, true / false positives & 9 / 0 & 9 / 0 & 9 / 4 & 9 / 0 \\
Shuffled sequences given any domain & 1/60 & 0/60 & 28/60 & 0/60 \\
\bottomrule
\end{tabular}
\end{table}

\subsection{Structure is predicted above chance for recently released proteins, and function is named correctly for half or fewer}

\begin{figure}[ht]
\centering
\includegraphics[width=\linewidth]{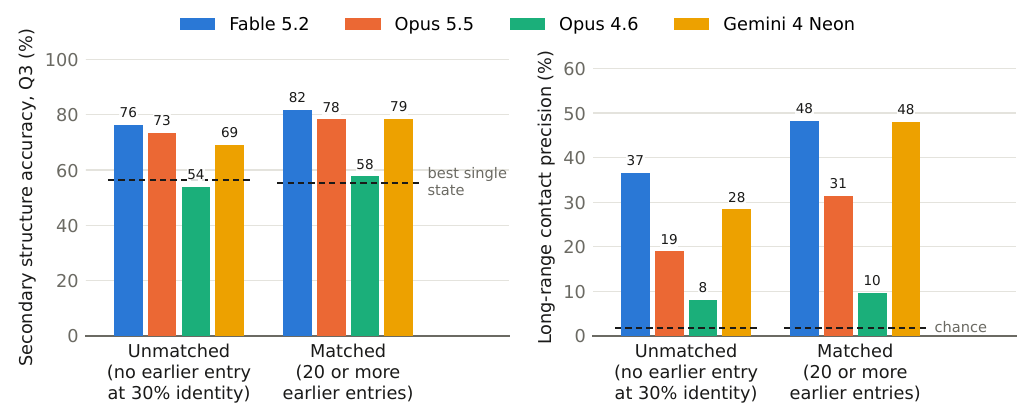}
\caption{Structure prediction from sequence for PDB entries released from August 2026, with ASPRV1
counted once. Unmatched: no PDB entity at 30\% identity or more released before June 2026. Matched: at least 20
such entities. There are 12 unmatched and 15 matched proteins; for Opus~5.5, the call for one matched entry
returned an API error instead of an answer. Claude models at high effort; Gemini~4 Neon at medium effort.
Dashed lines show the best single-state guess (left) and the chance rate for a long-range pair (right).}
\label{fig:struct}
\end{figure}

For the 12 unmatched proteins, Fable~5.2, Opus~5.5 and Neon reached 76\%, 73\% and 69\% three-state accuracy,
against 57\% for the best single-state guess. Their long-range contact precision was 37\%, 19\% and 28\%,
against 2\% by chance (Figure~\ref{fig:struct}, Table~\ref{tab:structure}). Precision here is over the
pairs that met the request, at least 24 residues apart with both residues resolved in the structure.
Counting every submitted pair, with a pair that did not qualify counted as wrong, gives 34\%, 18\% and 28\%. For the matched stratum,
accuracy was 78 to 82\% and contact precision 31 to 48\%. Twelve proteins are too few to rank two models
by a numerical difference.

Opus~4.6 did much worse. On the unmatched proteins, at high effort, it reached 54\% on
secondary structure, below the single-state guess, and 8\% contact precision, above chance but far below
the first three. At low effort it reached 48\% and 4\%.

We did not run dedicated structure predictors such as AlphaFold \citep{jumper2021alphafold} on these
proteins, so we make no comparison with them. We claim only that three models do well above chance without an alignment and with built-in tools turned off. Model class,
effort and route vary, so this is not a controlled comparison between generations or providers.

We also asked each model whether it recognised the protein. Of the 12 unmatched proteins, Fable~5.2 said it
did for 2 (ASPRV1 and MrtR), Neon for 1, Opus~5.5 for none and Opus~4.6 for 4. These self-reports do not
track accuracy: Opus~4.6 claimed the most recognitions and had the lowest structure scores. Recognition counts therefore cannot establish
recall or training exposure, and a wrong structure does not by itself prove a recognition claim false.

\begin{table}[ht]
\centering\small
\caption{Structure scores (percent), with ASPRV1 counted once: 12 unmatched and 15 matched
proteins, except 14 matched for Opus~5.5. Q3 is three-state accuracy; contacts are mean per-protein
long-range precision over the submitted pairs that were at least 24 residues apart with both residues
resolved. Effort labels are not comparable reasoning budgets. The Claude rows retain calls
listing more than one served model.}
\label{tab:structure}
\begin{tabular}{llrrrr}
\toprule
 & & \multicolumn{2}{c}{Unmatched} & \multicolumn{2}{c}{Matched} \\
\cmidrule(lr){3-4}\cmidrule(lr){5-6}
Model & Effort & Q3 & Contacts & Q3 & Contacts \\
\midrule
Claude Fable 5.2 & high & 76.3 & 36.6 & 81.8 & 48.3 \\
Claude Opus 5.5 & high & 73.4 & 19.0 & 78.3 & 31.4 \\
Gemini 4 Neon & medium & 69.0 & 28.4 & 78.6 & 48.0 \\
Claude Opus 4.6 & high & 53.9 & 8.1 & 57.8 & 9.7 \\
Claude Opus 4.6 & low & 48.1 & 4.1 & 45.1 & 4.0 \\
\bottomrule
\end{tabular}
\end{table}

Two graders, working separately and without knowing which model wrote an answer, graded the function
calls of the three frontier models for the 12 unmatched proteins against the deposited title. Both graders
were instances of a Claude model, not people (Methods). They agreed on 32 of the 36 overall grades. Fable~5.2
was correct on 5 proteins and partly correct on 2, Opus~5.5 on 4 and 1, and Neon on 2 and 1. The rest
were wrong or were abstentions: 2 or 3 wrong calls each for Fable~5.2 and Opus~5.5, depending on the
grader, and 4 to 6 for Neon. The four disagreements were all over whether a vague answer was an
abstention or an error. No model was correct on more than half of the proteins. Most of the correct
calls were for members of large, well-characterised families. We did not test whether the models got
there through the predicted structure or by recognising the family from sequence.

Table~\ref{tab:perprotein} gives the three measures for each protein. They do not move together. Fable~5.2
placed 18 of 19 contacts correctly for the kinase domain 9ZWQ and none for 9TGX or 9YP0, with similar
secondary-structure accuracy on all three. Secondary structure, tertiary contacts and function are
separate accomplishments.

\begin{table}[ht]
\centering\small
\caption{The 12 unmatched proteins, one by one. Q3 in percent, with the best single-state guess under
``Base''. Contacts are correct pairs over qualifying pairs. Function is the blinded overall grade:
C correct, P partly correct, W wrong, A abstained; two letters where the graders differed. Names are shortened
from the deposited entries.}
\label{tab:perprotein}
\footnotesize\setlength{\tabcolsep}{3.5pt}
\begin{tabular}{llr rll rll rll}
\toprule
 & & & \multicolumn{3}{c}{Fable 5.2} & \multicolumn{3}{c}{Opus 5.5} & \multicolumn{3}{c}{Gemini 4 Neon} \\
\cmidrule(lr){4-6}\cmidrule(lr){7-9}\cmidrule(lr){10-12}
PDB & Protein & Base & Q3 & Cont. & Fn & Q3 & Cont. & Fn & Q3 & Cont. & Fn \\
\midrule
37QB & Ketopantoate reductase PanG & 63 & 84 & 12/15 & P & 81 & 12/20 & P & 77 & 9/20 & W \\
9TGX & Guanidine prenyltransferase NosH & 86 & 82 & 0/19 & W & 84 & 0/7 & A & 76 & 0/20 & A \\
9YP0 & Designed cysteine protease & 43 & 79 & 0/20 & W & 76 & 0/16 & W & 72 & 0/20 & A/W \\
10ZI & Regulator MrtR & 65 & 79 & 5/20 & P & 76 & 4/20 & C & 79 & 9/20 & C \\
23AA & Anti-CRISPR AcrIIA17 & 58 & 78 & 2/19 & A & 73 & 1/19 & A & 72 & 0/20 & A \\
9SA5 & CBASS nucleotide cyclase & 54 & 78 & 6/20 & C & 71 & 5/19 & C & 59 & 4/20 & A/W \\
9ZWQ & N4BP2 polynucleotide kinase & 53 & 76 & 18/19 & C & 72 & 8/20 & C & 70 & 14/20 & C \\
27IK & ASPRV1 aspartic protease & 45 & 75 & 15/20 & C & 69 & 4/16 & C & 63 & 17/20 & P \\
9T71 & AtDUC1 deubiquitinase domain & 39 & 75 & 2/19 & C & 70 & 0/19 & A & 61 & 1/17 & A \\
9WIA & Insect effector HARP1 & 57 & 75 & 4/19 & A & 64 & 2/16 & W & 71 & 1/20 & W \\
12WN & Designed protein desE2D & 60 & 72 & 11/20 & C & 80 & 6/18 & A & 75 & 9/20 & W \\
10CW & RNase P subunit Rpp30 & 56 & 62 & 7/19 & W/A & 65 & 1/18 & W/A & 54 & 4/20 & W \\
\bottomrule
\end{tabular}
\end{table}

For a domain of \textit{Arabidopsis} AtDUC1 (PDB 9T71), deposited under the name ``EEIG1/EHBP1 protein
amino-terminal domain protein'', Fable~5.2 predicted a papain-like cysteine peptidase with a Cys-His-Asp
triad, ``possibly a deubiquitinase''. The entry is the structure of its deubiquitinase domain. This call
cannot be credited to the sequence alone, however. A preprint describing this protein family as
deubiquitinases was posted in December 2025 \citep{lachmayer2025ducs}, eight months before the structure
was released, and a model trained after that date could have learned it. We did not search for earlier
descriptions of the other unmatched proteins.

Good secondary structure did not imply a correct function. For the guanidine prenyltransferase NosH
(9TGX), Fable~5.2 scored 82\% three-state accuracy on an almost entirely helical protein, where calling
every residue helix scores 87\%. None of its 19 qualifying contacts was right, and it named the protein a
TenA-family thiaminase. No model identified the RNase~P subunit Rpp30 (10CW). For a designed cysteine
protease (9YP0), Fable~5.2 and Opus~5.5 each described a natural bacterial surface protein with no
catalytic function. Both graders counted these as wrong claims and not as abstentions.

Neon's calls show the same gap between recognising a
family and annotating a protein. Neon correctly named a polynucleotide kinase for 9ZWQ and an
AHL-responsive LuxR-family regulator for 10ZI. For the retained ASPRV1 entry (27IK), it predicted the
right family, the aspartic proteases, but named the protein DDI2 and gave it DDI2-specific substrate
biology. It also assigned PanG (37QB) to the TyrA dehydrogenase family and Rpp30 (10CW) to a metabolic
enzyme.

One of the ASPRV1 constructs (27IQ) gave an unplanned repeat of the \tpseudo{} test, because its catalytic DSG motif
is changed to NSG. Without being asked about activity, Fable~5.2 and Opus~5.5 said that the catalytic
aspartate was replaced and that the construct was probably inactive. Neon did not identify the
substitution.

\subsection{A next-residue advantage is detectable for two of the four models tested}\label{sec:nextres}

The report that prompted this study measured next-residue prediction, and we repeated that measurement on
a subsample of its questions. Fable~5.2 predicted real sequence better than shuffled sequence at both effort settings, and Opus~5.5 did
so at high effort. At low effort, the interval for Opus~5.5 includes zero. Opus~4.6 and Gemini~4 Neon,
each run at low effort only, showed no difference (Table~\ref{tab:next}). Neon scored 8.8\% on real and
10.4\% on shuffled sequence, and none of its ten-residue continuations matched exactly. Accuracy was well
below the published values for the identical questions (58.8\% for Fable~5.2 and about 40\% for
Opus~5.5). We suspect the cause is sampled answers and the extra context our calls carried, but we did
not test this. Reversed sequence, tested for the Claude models at low effort only, scored like shuffled
sequence for Fable~5.2 and between shuffled and real for Opus~5.5. Four simple heuristics that use no
model, such as repeating the last residue, scored 7 to 10\% on real, shuffled and reversed sequence alike
over all 3,340 cut points.

Accuracy on real sequence rose with the effort setting. On the same 136 real cut points, Fable~5.2 scored
25.0\% at low effort and 41.2\% at high effort, where every answer used thinking tokens. On shuffled
sequence it scored 7.4\% and 11.8\%. This is the only controlled evidence on reasoning in this study. It
differs from the original report, in which answers without thinking were as accurate as those with it.

Within a run, answers that used thinking were more often right. The model chooses when to think, though,
so this comparison describes the runs and does not measure what thinking does. At low effort, Fable~5.2
thought on 38 of 136 real cut points and was correct on 57.9\% of them, against 12.2\% of the other 98.
It thought on only 15 shuffled and 17 reversed cut points. On shuffled sequence, thinking did not help:
0 of 15 were correct with it, against 10 of 121 without. At high effort, where it thought on every
question, its accuracy on real sequence was 41.2\%, well below the 57.9\% on the questions it had picked
itself. The model seems to think when it recognises something, so the split partly sorts questions by
difficulty. Opus~5.5 at high effort was correct on 37.2\% of the 43 real cut points where it thought and
on 10.8\% of the 93 where it did not. We cannot see the reasoning text.

\begin{table}[ht]
\centering\small
\caption{Next-residue accuracy on a 68-family subsample. Gap is real minus shuffled, in percentage points,
with a 95\% bootstrap interval over families. Neon answered 136 real and 135 shuffled questions and used
thinking on every one; Fable~5.2 at low effort did so on 38 of 136 real cut points. Effort labels are not
comparable across providers.}
\label{tab:next}
\begin{tabular}{llrrrl}
\toprule
Model & Effort & Real & Shuffled & Reversed & Gap (95\% interval) \\
\midrule
Fable 5.2 & low & 25.0\% & 7.4\% & 8.8\% & $+17.6$ ($+9.6$ to $+26.5$) \\
Fable 5.2 & high & 41.2\% & 11.8\% & -- & $+29.4$ ($+18.4$ to $+40.4$) \\
Opus 5.5 & low & 16.9\% & 9.6\% & 12.5\% & $+7.4$ ($-1.5$ to $+16.2$) \\
Opus 5.5 & high & 19.1\% & 7.4\% & -- & $+11.8$ ($+3.7$ to $+20.6$) \\
Opus 4.6 & low & 5.1\% & 7.4\% & 5.9\% & $-2.2$ ($-7.4$ to $+2.9$) \\
Gemini 4 Neon & low & 8.8\% & 10.4\% & -- & $-1.5$ ($-8.1$ to $+5.1$) \\
\bottomrule
\end{tabular}
\end{table}

\begin{table}[ht]
\centering\small
\caption{Median recorded thinking tokens per scored answer, zeros included; no scored row lacks a count.
The \tnr{} test includes real and shuffled questions only, and the \tfeat{} test both conditions. The Opus~5.5 structure
median is over 29 answers, with its API error excluded.}
\label{tab:thinking}
\begin{tabular}{lrrrr}
\toprule
Model & \tpseudo & \tfeat & \tstrfn & \tnr \\
\midrule
Gemini 4 Neon & 2,474 & 1,049 & 17,881.5 & 634 \\
Claude Fable 5.2 & 2,548 & 703.5 & 3,009 & 164 \\
Claude Opus 5.5 & 1,275.5 & 418 & 1,342 & 0 \\
Claude Opus 4.6 & 19,921 & 14,189.5 & 42,987 & 0 \\
\bottomrule
\end{tabular}

\smallskip
\parbox{\linewidth}{\footnotesize Effort: Neon low in the \tnr, \tfeat{} and \tpseudo{} tests and medium in the \tstrfn{} test; Fable and
Opus~5.5 high throughout; Opus~4.6 low in the \tnr{} test and high in the others. For the \tfeat{} test, the Fable and
Opus~5.5 setting comes from the authors' record of the collection, since those rows do not store it.
The counts do not show that budgets were matched.}
\end{table}

\FloatBarrier
\section{Discussion}

Three of the four models, Fable~5.2, Opus~5.5 and Gemini~4 Neon, do more than recognise a family. When one
catalytic residue changes, they change their answer and name that residue. They predict secondary
structure and contacts above chance for proteins with no earlier PDB entry at 30\% identity. And they
rarely assign a Pfam domain to a shuffled sequence. The three come from two providers and were reached by different
routes, so the behaviour is not peculiar to one vendor or one way of calling a model. We have not
established how far it extends.

Opus~4.6 shows that the behaviour is not general. It found all 9 signal peptides, 67
of 74 transmembrane helices and the kinase aspartate substitution. But it missed the lysine substitution
and the active-site changes in other enzyme families, and it invented a domain for nearly half of the
shuffled sequences. This contrast only describes what we
saw. Effort labels do not order the thinking actually recorded (Table~\ref{tab:thinking}), access routes
differ, and no model was run at every setting.

For Fable~5.2 and Opus~5.5, the decoy control indicates that the calls track the catalytic position.
Because the decoys avoided other conserved positions, the control does not show how the models treat
changes there, and it was not run for Neon. In two cases where the models disagreed with the curated
site, alignment and structural evidence favours the positions they proposed, more strongly for YrfG than
for ENV9. A sequence-only read may therefore also help flag sites that were misplaced when annotation was
transferred by similarity. Two cases do not give a rate, though.

The other abilities appeared without a detectable advantage in our next-residue test. Neon showed no
detectable next-residue advantage, yet it rejected all 20 aspartate mutants, reported 16 of 20 lysine
substitutions and reached 28.4\% contact precision against 1.8\% by chance. The tasks dissociate, but
that does not prove separate mechanisms or an absence of sequence knowledge, and a null result is weak
evidence. Neon was run on the \tnr{} test at low effort only. Opus~5.5 at low effort also showed no detectable
advantage. Neon used thinking on every question it answered and Opus~5.5 at low effort on none, so we
cannot tell whether Neon's null result comes from the model or from the setting. The Claude models
scored well below the published values on the identical questions (Section~\ref{sec:nextres}), so this
setup depresses the test for every model. On its main test set, the cited report ran Gemini~4 Carbon
and, by raw continuation, Gemini~2.5 Pro, but not Neon. It used temperature 0 where accepted and
thinking off or at the lowest setting, and found both models near zero \citep{tu2026claude}. Neon's
result is consistent with theirs but does not replicate it.

For annotation review, the practical use is narrow. Before a catalytic term is accepted, a sequence-only
read can flag a protein that has lost a canonical catalytic residue, and it can question an implausible
name. It cannot show that the protein is inactive, since catalytic residues can be displaced
\citep{goldberg2023pseudoenzymes}. Nor can it find a protein that keeps every catalytic residue and
still shows no activity, as reported for kin-32. A flag should prompt a curator to check the catalytic term. It says nothing against the
protein having a function, since dead enzymes commonly act as regulators \citep{adrain2020dead}.

By itself the read is also not enough to support a curated annotation. Its output has no provenance, its errors are stated as
confidently as its correct calls, and both patterns we observed among the mutants called active end in a
confident ``active''. Any
catalytic position a model names should be checked against the sequence and a curated feature before it
is used.

\paragraph{A neurosymbolic outlook.} Neurosymbolic AI combines sub-symbolic learning, which finds patterns
in raw data, with symbolic knowledge and reasoning \citep{garcez2023neurosymbolic,kautz2022third}. Protein
annotation has long divided the work in this way. Profile models and protein language models do the
sub-symbolic part: they find a domain or a motif in a string of residues, which a person cannot do by
reading it. Curators and annotation rules do the symbolic part: a kinase that has lost the aspartate of
its catalytic loop is unlikely to be active, and a catalytic term should not be transferred to it. In
current practice the two parts are joined by pipelines and by people.

In our tests, the three stronger models did both parts in a single answer. They placed Pfam domains in
raw sequence, which is sub-symbolic work, and they reasoned about what a substituted residue does, which
is symbolic. The aspartate mutants show the combination. Given
the rhoptry kinase ROP17 (Q2PAY3) with Asp436 changed to asparagine, Fable~5.2 found the glycine loop,
the $\beta$3 lysine and the DFG aspartate intact, noted that the catalytic-loop HRD motif now read HGN,
and concluded that the loss of the catalytic base made the protein a pseudokinase. Stated
reasons need not reflect how a model reaches its answer. The decoy control does not depend on them,
though: the calls of Fable~5.2 and Opus~5.5 tracked the catalytic position whatever explanation came
with them.

This suggests a different design for annotation review. A single model could read the sequence, recall
what the literature and curated rules say about the family, and weigh experimental data, then report
where these disagree, as the models did for YrfG. The sequence-reading part has not been compared with
dedicated predictors, and its output has no provenance, so in practice the model would propose and
tools would check: an alignment to reviewed relatives, a predicted structure, the curated feature. Our
tests do not show that such a combination improves annotation. They show that the three frontier
models we tested already have both of the parts it would need.

\paragraph{Limitations.} The tests are small. ``Frontier'' here means three models from two providers, and
the contrast with earlier models rests on one, Opus~4.6. Cells contain 12 to 68 items, and each question has at most
one scored answer per model, effort setting and run. For the Claude models, the \tpseudo{} mutants and wild types were asked again at the same effort for
the decoy control. Calls marked as carrying attached context, or whose attempt failed, were also asked
again, and the marked rows are not scored. The split of \tnr{} accuracy by thinking is observational.

Settings and routes were not matched. Answers were sampled. Claude questions went through a command-line
tool that adds its own context, and Gemini questions through provider APIs. We measured neither the
effect of the route nor any benefit of greedy decoding, and effort settings are not matched across
providers. The Opus~4.6 high-effort runs used a revised call route and a later tool version than the runs
of the other two Claude models.

The controls of the \tpseudo{} test have gaps. The test shows that calls differ between substitutions
at annotated catalytic positions and matched substitutions elsewhere. It does not show that any call is
correct: we assumed, and did not test, that the catalytic-site mutants are inactive and that the decoys
are neutral. No decoy control was run for Neon, and the Opus~4.6 decoys
were asked at low effort only. The decoy positions were chosen by a rule written with the help of two of
the tested models, and they avoid conserved positions. The adjudication of the two reference sites rests
on alignment and predicted structures. Neither disputed position, nor the activity of the mutants used
here, has been tested directly by experiment. Swiss-Prot proteins are well known, so the wild-type
results in the \tpseudo{} test may reflect recall. The mutants are new sequences, so a changed call shows more than a repeated label. A model could
still recognise the parent protein and recall which residues its catalysis needs. Mutants made by one
substitution are also not natural pseudoenzymes. Those usually differ from active relatives at many
positions and can be inactive for reasons that leave the catalytic residues intact. We examined no
over-annotated entry.

The reference data and grading have limits too. Transmembrane and signal-peptide truth in the \tfeat{} test is
itself predicted. Function calls in the \tstrfn{} test were graded by two blinded instances of a Claude model, the
family of two of the three models graded, and by no human. They were graded against the deposited title alone, for
12 proteins. The unmatched
stratum of the \tstrfn{} test excludes close homologs in the PDB and nothing else. Most of its sequences were already
public, remote homologs were not excluded, and at least one protein was described in a preprint before
its structure was released. The random-contact baseline does not substitute for running dedicated
predictors on the same proteins.

We did not test eukaryotic multi-domain proteins in the \tfeat{} test, membrane proteins in the \tstrfn{} test or general
variant-effect prediction, which PG-LLM benchmarks \citep{arora2026pgllm}. Apart from Opus~4.6, we did not test whether more reasoning effort improves results
in the \tfeat, \tpseudo{} and \tstrfn{} tests. Finally, we have no information on how these models were trained and draw no conclusion
about it.

\FloatBarrier
\section{Methods}\label{sec:methods}

\paragraph{Claude calls.} Each question was asked through the \texttt{claude} command-line tool in print
mode, with its built-in tools turned off and an empty system prompt. The scripts of 18 November retried a
question up to three times when the tool's output was not valid JSON. They kept only the last attempt and
recorded none of the earlier ones. The later runs record the number of attempts in each row, and in those
runs a call marked as carrying attached context was kept as a row and asked again. The call counts below
are counts of recorded rows.

This route does not expose temperature, so answers were sampled. It also adds context of its own, whose
size the rows of 18 November do not record. With the original flags a one-sentence prompt carried 5,851 input tokens, and with the revised flags 573 (Supplementary Methods).

Fable~5.2 and Opus~5.5 ran at high effort in the \tfeat, \tpseudo{} and \tstrfn{} tests; the rows of the \tfeat{} runs of 18 November do
not record the effort setting. Opus~4.6 was asked the \tfeat, \tpseudo{} and \tstrfn{} tests twice: at low effort on 18 November by the
original route, and at high effort on 19 November by the revised route. Unless stated, we report the
high-effort run, which matches the other two models in effort but not in route.

Each question has at most one scored answer per run. For 6 of the 2,854 recorded calls in the four tests,
the tool's usage record lists a second model besides the one requested. One of them returned an API
error and is excluded. The other five are kept, and excluding them changes no conclusion (Supplementary
Methods).

\paragraph{Gemini calls.} Gemini~4 Neon was asked on 19 November 2026 through
the global Vertex AI \texttt{generateContent} endpoint, with no tools or system instruction in the request
body. Temperature was not specified.
The \tnr{} test used real and shuffled sequences only. Neon gave usable answers for 271 of 272 questions in the \tnr{} test
(one shuffled question was safety-blocked), all 120 in each of the \tfeat{} and \tpseudo{} tests, and all 30 in the \tstrfn{} test. Successful
API responses finished with \texttt{STOP}. Retry rules, and a sensitivity analysis for six \tnr{} questions with
unexplained extra input, are in the Supplementary Methods.

\paragraph{Recorded thinking.} Rather than infer a budget from the effort label,
Table~\ref{tab:thinking} reports the median recorded thinking tokens per scored answer, zeros included.
On structure, Neon used a median of 17,881.5 thinking tokens at medium effort, compared with 3,009 for
Fable and 1,342 for Opus~5.5 at high effort. The counts come from different tokenizers and reporting
routes and are not equal units of compute, and retries and continuations may be counted differently.
None of the completed Neon runs used its highest effort setting, so its best achievable performance is
unmeasured. These single-run scores are not a statistical lower bound on it either, because higher
effort need not improve every task; the Opus~4.6 shuffled-domain results are an example.

\paragraph{Catalytic-site mutants and proteins named inactive (\tpseudo).} A catalytic-site mutant here
is an active enzyme in which we substituted one annotated catalytic residue. We expect such a mutant to be
inactive, but we did not test it, and mutation of a catalytic residue is not always a good indication of
lost activity \citep{ribeiro2020pitfalls}. The endpoint of the test is therefore the model's sensitivity
to the substitution, and inactivity is a prediction. A natural pseudoenzyme is an evolved homologue, which
usually differs from active relatives at many positions. UniProt curates pseudoenzymes in Swiss-Prot
\citep{zaru2020uniprot}. In UniProt release \texttt{2026\_04}, queried on 19 November 2026 (UTC), 688
reviewed entries carried ``inactive'' in a protein name (492 in the recommended name), 107 of them in the
protein kinase superfamily. From Swiss-Prot, excluding human, mouse and rat, we took three sets of 20:
active protein kinases with a curated proton-acceptor aspartate and ATP-binding lysine; kinase-fold
proteins whose name contains ``inactive''; and non-kinase enzymes with a curated active site. The second
set is defined by the Swiss-Prot name alone. We did not curate those 20 labels, and a name of this kind
can rest on a missing residue or on one assay. Each kinase
was also presented with the aspartate changed to asparagine and, separately, with the lysine changed to
methionine. Each other enzyme was presented with its first active-site residue changed to alanine. A
single prompt, used for every sequence, asked for the catalytic residues present, any that were
substituted or missing, and whether the protein is catalytically active.

\paragraph{Decoy control for the \tpseudo{} test.} A model that calls any altered sequence inactive would pass the \tpseudo{} test
without reading the catalytic residues. For each catalytic-site mutant we therefore made a decoy, with the
same substitution type in the same protein at a position with no functional annotation. The absence of
an annotation does not show that the substitution is neutral. Kinase decoys lie
inside the kinase domain. We excluded positions within three residues of any annotated site and, in
kinases, the catalytic loop, the DFG motif and the subdomain~IX aspartate, then drew the position with a
fixed seed. Two enzymes had no eligible residue, which left 58 decoys in 38 proteins, a median of 46
residues from the nearest curated catalytic position. Decoys were asked with the \tpseudo{} prompt by the
revised route. The wild types and catalytic-site mutants of the same proteins were asked again by that
route, so that all three are compared within one route. The control was run for the three Claude models
only: Fable~5.2 and Opus~5.5 at high effort, and Opus~4.6 at low effort. It added 174 decoy calls and 300
repeat calls. The design does not test a substitution at another conserved position.

\paragraph{Adjudication of two reference sites.} We selected two enzymes with disputed catalytic
positions after inspecting the model answers. We aligned each with reviewed UniProt relatives carrying an
experimentally supported active site (ECO:0000269), using MAFFT \citep{katoh2013mafft}. For ENV9 we also
included its cited annotation source, O93868, and aligned the pair separately. We measured distances to
other catalytic residues and to any available metal or cofactor atoms in SWISS-MODEL homology models
\citep{waterhouse2018swissmodel} and in AlphaFold models with ligands transplanted by AlphaFill
\citep{jumper2021alphafold,hekkelman2023alphafill}. These predictions share homology information, and the
SWISS-MODEL template identities are only 14.16\% (YrfG) and 18.48\% (ENV9). No answer from a tested model
is used as evidence. Excluding these cases is a post hoc sensitivity analysis, and the main scores keep
them.

\paragraph{Domains and membrane features (\tfeat).} The test set was 45 unreviewed bacterial and archaeal
UniProt entries of 150 to 450 residues with a Pfam match, drawn haphazardly through the search interface
\citep{uniprot}, and 15 proteins from the test set of Tu and Chen \citep{tu2026probe}. The model returned
Pfam domains with coordinates \citep{mistry2021pfam}, transmembrane helices, a signal-peptide call, a name
and a function. Truth was the InterPro Pfam matches and the UniProt transmembrane and signal features,
which in unreviewed entries are themselves predicted. Each protein was also presented shuffled.

\paragraph{Structure and function (\tstrfn).} From the PDB \citep{rcsb} we took X-ray entries released on
or after 1 August 2026 with one protein entity of 100 to 350 residues and a resolution of 2.5~\AA{} or
better. We kept 15 entries for which a sequence search found no PDB entity at 30\% identity or more
released before 1 June 2026, and 15 entries with at least 20 such entities. We call these strata
``unmatched'' and ``matched'' for short, and the labels describe that criterion only (the data files call
them \texttt{novel} and \texttt{known}). Unmatched entries may have
more distant homologs in the PDB. Of the 12 distinct unmatched proteins, 8 have UniProt accessions, so their
sequences were public earlier. Structures are deposited months before release, and the work behind an
entry may have been described in a preprint. We do not know the models' training cutoffs. The two dates
rest on our assumption that structures released from August 2026 were not in training, and the date and
homology filters do not establish it.

The model predicted helix and strand ranges, a fold, 20 residue pairs at least 24 apart in sequence that
it expected to lie within 8~\AA{} (C$\alpha$), a function, and whether it recognised the sequence as a
specific known protein. Truth was the deposited coordinates and the depositors' helix and sheet records,
scored over resolved residues. Contact precision is computed per protein over the submitted pairs that
were at least 24 apart with both residues resolved, and then averaged. The models did not always comply:
of 240 pairs submitted for the 12 unmatched proteins, 229 qualified for Fable~5.2 and 208 for Opus~5.5. We
therefore also report precision over all submitted pairs. One protein, human ASPRV1, was drawn four times
in the unmatched set, and we count it once (12 unmatched proteins).

Function calls for these 12 proteins were graded twice. The AI agent that ran the Claude tests first
graded Fable~5.2 and Opus~5.5 once, unblinded, as correct, abstaining or wrong. That grading counted
confident descriptions of the wrong kind of protein as abstentions, so we replaced it. For the grades
reported here, the answers of Fable~5.2, Opus~5.5 and Neon were stripped of model names and shuffled. Two
separate instances of a Claude model each graded all 36 against the deposited title and molecule name.
Each recorded whether the family, the specific function and any claim about the protein's identity or
origin were right, whether the answer was a true abstention, and an overall grade (correct, partly
correct, wrong or abstained). The graders agreed on 32 of 36 overall grades (Cohen's $\kappa = 0.85$), and
we give both counts where they differ. Both gradings are in the repository.

\paragraph{Next residue (\tnr).} We drew 68 of the 167 families in the public test set
\citep{tu2026probe}, 17 per family-size bin, and took one protein per family and two of its five cut
points. We used the authors' prompt unchanged. The model writes the next ten residues, and the first is
scored. The controls were the same protein shuffled with the authors' seed and, new here, the real
sequence reversed. Intervals are 95\% bootstrap intervals over families.

\section*{Data and code availability}

Scripts, prompts, test sets and all raw model outputs are at
\url{https://github.com/ai4curation/fable-sequence-to-function}, and a summary of the results is at
\url{https://ai4curation.io/fable-sequence-to-function/}. Scoring scripts print the quantitative
summaries in this paper from the saved model outputs, and \texttt{just check} verifies the data checksums
and that the numbers reproduce. Test sets were drawn from live UniProt, InterPro and RCSB services on 18
November 2026. They are committed as drawn, because re-querying those services returns different entries.
By the authors' record, the Claude calls of 18 November used Claude Code 2.1.327; those rows do not store
the version. The eight Opus~5.5 next-residue questions asked on 19 November, the Opus~4.6 high-effort
runs and the decoy control used 2.1.328, as their rows record. The reported Gemini calls went to the
Vertex AI endpoint. Repeating either set of calls will not return identical answers.

The Gemini collection scripts, raw replies, incomplete pilots and an offline audit of prompts and
coverage are in \texttt{gemini\_experiment/}. These data are included in the checksums and frozen
reports.

\section*{Acknowledgements}

This work was supported by the U.S. Department of Energy, Office of Science, under Contract No.
DE-AC02-05CH11231 to Lawrence Berkeley National Laboratory.

\bibliographystyle{unsrtnat}
\bibliography{references}

\clearpage
\appendix
\section*{Supplementary Methods}\label{sec:supp}

\paragraph{Context carried by Claude calls.} We measured it on 19 November with the same context flags and
Claude Code 2.1.328. A one-sentence prompt to Haiku~5.5 carried 5,851 input tokens. Two questions with the
\tpseudo{} prompt to Fable~5.2 carried 97,813 and 6,413; the larger one most likely had the account's
connector tools attached. We do not know whether the stored calls carried such an attachment. Their rows
record cost but not tokens, and the tool caches repeated context, so a cached attachment would add far
less to the cost of each later call. Eight Opus~5.5 next-residue questions at high effort were left
unasked when that run was interrupted. We asked them on 19 November (UTC) with Claude Code 2.1.328 and
flags that keep user settings and connector tools out of the call. With these flags, the one-sentence
prompt carried 573 input tokens.

\paragraph{Continued replies.} The high-effort Opus~4.6 answers were long. The tool asks again by itself after an output limit or a stopped reply, and 7 of 120 replies in the \tpseudo{} test
and 9 of 30 in the \tstrfn{} test took more than one request for this reason. In the \tfeat{} test, six such replies were asked
again, and one of the new answers also took more than one request. In the \tnr{} test, Opus~4.6 was run at low
effort only.

\paragraph{Web requests.} Counts of web search and fetch requests were kept only in rows written by later versions of the revised
route's code. These are mostly the Opus~4.6 high-effort runs of the \tpseudo{} and \tstrfn{} tests and the rerun of the \tpseudo{}
mutants and wild types for the decoy control. None of those rows shows a web request.

\paragraph{Calls listing a second model, and replies with more than one answer.} Of the six calls whose usage record lists a second model, three are Opus~5.5 next-residue
calls at high effort, all on shuffled sequences. The other three are in the \tstrfn{} test, one Fable~5.2 and two
Opus~5.5 calls, all on matched entries. One of the Opus~5.5 \tstrfn{} calls (PDB 7IMP) returned an
API error instead of an answer and is excluded from the scores. For the other five, the record does not
show which model wrote the answer. We kept them in the reported numbers; the repository also gives the \tnr{} and \tstrfn{} tests with them excluded, and no conclusion changes. Where a reply held more than one JSON object, we
scored the last one. This applies to three replies of the earlier Opus~4.6 run of the \tfeat{} test, in which the
second object corrects the first. One Opus~5.5 next-residue reply was a written-out web request instead
of a prediction; it is counted as invalid and as a wrong answer.

\paragraph{Gemini retries and input sizes.} Requests that met a rate limit, a server error, a Vertex authorisation error or a client exception were
sent again, up to a fixed number of tries; safety blocks were not. Attempt counts are retained, but not
all intermediate responses or their token usage. The retry rule changed during collection. Whether an earlier version also re-asked after
a completed but unwanted reply cannot be checked from the files. For six Neon \tnr{} questions, the
endpoint recorded 764 to 921 input tokens, or 1.3 to 1.9 tokens per prompt character, against at most
0.5 for the other 265. We could not establish what the extra input was. Excluding those six rows gives
8.3\% real accuracy (133 questions) and 10.6\% shuffled accuracy (132), with a family-averaged gap of
$-2.3$ points (95\% interval $-9.1$ to $+4.5$, 66 families), so the conclusion is unchanged. Incomplete
CLI and API pilots, including a four-question Gemini~4 Carbon structure run, are kept in the repository
and excluded from full-test comparisons. The two CLI pilots carried median inputs of about 9,800 and
9,900 tokens per call.
\end{document}